\documentclass[aps,prd,preprint,a4paper]{revtex4}
\usepackage{slashed}
\usepackage{epsfig}

\begin{document}

\title{On the quantum loop suppressed electroweak processes}

\author{Davor Palle}
\affiliation{
Zavod za teorijsku fiziku, Institut Rugjer Bo\v skovi\' c \\
Bijeni\v cka cesta 54, 10000 Zagreb, Croatia}

\date{May 15, 2013}

\begin{abstract}
The quantum loop suppressed electroweak processes appear
to be very sensitive probes for the symmetry-breaking mechanisms.
Since the Standard Model does not involve massive neutrinos,
baryon or lepton number violations and the cold dark matter
particle, there are great expectations in search for physics
beyond the Standard Model. The LHC experiments reported no new
physics, except for the appearance of the 125 GeV heavy resonance. 
If this resonance turns out to be a Higgs boson, we shall be
faced with the paradox that the Standard Model, as a theoretically
and experimentally incomplete and defective theory, is confirmed.
If, on the other hand,
this resonance is a scalar or pseudoscalar meson -
a mixture of a (pseudo)scalar glueball and toponium, 
one has to study alternative symmetry-breaking mechanisms. This
paper is devoted to the evaluations of the gauge invariant electroweak
functions that are relevant for the suppressed electroweak processes within
the SM or the UV finite theory in which the noncontractible space breaks
the gauge symmetry.
Deviations from the SM amplitudes
range from 10\% to 30\%, thus being measurable at the LHCb and future
superB factories. The most recent result of the LHCb, measuring
certain B meson branching fraction, appears to be lower than the SM prediction,
thus favouring our theory of noncontractible space. 
\end{abstract}

\maketitle 

\section{Introduction and motivation}

The physics of suppressed electroweak (EW) processes by quantum
loops figure as the best window for search into the new
physics territory.
More than three decades of the theoretical work strengthens
our ability to predict their amplitudes with high precision.
While electroweak and hard QCD calculations rely on the
perturbative framework and the renormalization theory,
the hadron matrix elements are most accurately evaluated within the lattice
QCD.

The extensive search for new physics at the LHC reveals a
new boson particle \cite{CMS} with a mass around 125 GeV. 
Since the Standard Model (SM) does not contain massive neutrinos,
lepton or baryon number violations and a candidate particle
for the cold dark matter, it is very unlikely that the new
particle is the SM Higgs boson. However, any extension of the SM that has
to fulfil above improvements, such as SUSY or GUT extension,
implicates the existence of more scalar particles. It is
well known that the Higgs mechanism does not solve the problem of
particle masses but only breaks the gauge symmetry,
i.e. the fermion masses are free
parameters.

The alternative explanation of the 125 GeV resonance is
its identification as the scalar or pseudoscalar meson, which is
a mixture of the (pseudo)scalar glueball and toponium, suggested by
P. Cea \cite{Cea}, thus leaving the problem of the symmetry-breaking
mechanism unsolved.

Anyhow, the new LHC data with the total and partial decay widths of the
new particle, as well as its mass, spin and parity, should
resolve this puzzle. The first LHC results \cite{CMS} point
to a surpuls of photons in the decay
of the 125 GeV resonance assumed to be a SM Higgs scalar,
as well as to the absence of any firm decay signal into fermions.

In this paper, we present complete one loop results for the gauge 
invariant functions appearing in the suppressed EW processes
within the SM and the BY theory \cite{Palle1}, the theory without the Higgs
scalar but with the Lorentz and gauge invariant UV cut-off.
The next chapter is devoted to the explicit deccription of
the calculations with details in the Appendix, while the
concluding section provides numerical estimates and 
points out the difference between the SM and the BY theory.

\section{Evaluation of the gauge invariant functions}

In this paper, we shall omit a detailed description of the BY
theory with its mathematical \cite{Palle1} and phenomenological 
advantages in particle physics and cosmology \cite{Palle2,Palle3,Palle4}.
The Lorentz and gauge invariant cut-off of the BY theory
is fixed by the Wick's theorem and trace anomaly \cite{Palle1} through
weak coupling and the weak gauge boson mass   
$\Lambda=\frac{\hbar}{c d}=\frac{2}{g}\frac{\pi}{\sqrt{6}}
M_{W} \simeq 326 GeV$ where g=weak coupling, $M_{W}$=weak gauge boson mass.
The noncontractible space cannot essentially affect the strong coupling
below the scale $\mu \simeq 200 GeV$ \cite{Palle2}, hence the impact
of $\Lambda$ on the hard QCD corrections to the
electroweak processes \cite{Buras1} is negligible. Thus, we focus
onto the dominant one loop electroweak contributions written
in the form of the gauge invariant functions $A,X,Y,Z,E$ \cite{Buchalla}. 
These functions appear in the meson mixings $M^{0}-\bar{M^{0}}$ and
meson decays $M \rightarrow l^{+}l^{-},\ M_{1} \rightarrow M_{2}\nu\bar{\nu},
\ M_{1} \rightarrow M_{2}l^{+}l^{-},\ M_{1} \rightarrow M_{2}\gamma$
\cite{Inami,Buras2,Buchalla}.

The Inami-Lim functions $S_{0}(x_{t})=A(x_{t})$ and $Y(x_{t})$ are 
treated and compared between the SM and BY theory in refs. \cite{Palle5}
and \cite{Palle6}.
We have to reevaluate the self-energy, vertex and box type diagrams for
Z bosons, photons and gluons in
figs. 1-4 of ref. \cite{Inami} in both the SM and the BY theory to
get the explicit forms for the gauge invariant $X,Y,Z,E$ functions.

We list now our results obtained in the 't Hooft-Feynman gauge 
for the $B,C,D,E$ functions \cite{Buchalla}
referring to figures 1-4 in ref. \cite{Inami}:

$B_{1}$ box diagram as a sum of four graphs of Fig. 2 (a),(b),(c),(d)
of ref. \cite{Inami}:

\begin{eqnarray}
B_{1}=M^{2}_{W}[\frac{1}{4}f_{1}(m_{l},m_{\alpha},M_{W})
(1+\frac{1}{4}\frac{m_{l}^{2}}{M^{2}_{W}}\frac{m_{\alpha}^{2}}{M^{2}_{W}})
-\frac{1}{2}\frac{m_{\alpha}^{2}}{M^{2}_{W}}m_{l}^{2}
g_{1}(m_{l},m_{\alpha},M_{W})], 
\end{eqnarray}

$m_{l}$ is the lepton mass, $m_{\alpha}$ is the quark mass and
$M_{W}$ the weak boson mass.
For the definitions and analytic expressions of $g_{1}$ and $f_{1}$ functions, see
the Appendix.

$B_{2}$ box diagram as a graph of Fig. 2 (a)
of ref. \cite{Inami}:

\begin{eqnarray}
B_{2}=-\frac{1}{4}M^{2}_{W}L(M_{W},m_{\alpha}), 
\end{eqnarray}

for the definition and analytic expression of the L function, see
the Appendix.

$C = \frac{1}{2} \Gamma_{Z}=\frac{1}{2}\sum_{i=(a)}^{(h)}\Gamma_{Z}^{i}$ 
vertex is given in Fig. 3 (a-h) of
ref. \cite{Inami} and is reevaluated in ref. \cite{Palle6}:

\begin{eqnarray*}
\Gamma_{Z}^{(a+b)}&=&(-\frac{1}{2}+\frac{1}{3}s_{W}^{2})(1+\frac{1}{2}
\frac{m_{j}^{2}}{M_{W}^{2}})\tilde{B}_{1}(0;m_{j},M_{W}), \hspace{50 mm} \\
\Gamma_{Z}^{(c)}&=& -\frac{1}{4}(-1+\frac{4}{3}s_{W}^{2})
(\tilde{B}_{0}(0;m_{j},M_{W})+m_{j}^{2}L(m_{j},M_{W}))+
\frac{2}{3}s_{W}^{2}m_{j}^{2}L(m_{j},M_{W}),  \\
\Gamma_{Z}^{(d)}&=& -\frac{1}{2}\frac{m_{j}^{2}}{M_{W}^{2}}
(\frac{1}{3}s_{W}^{2}\tilde{B}_{0}(0;m_{j},M_{W})
+m_{j}^{2}(\frac{1}{2}-\frac{1}{3}s_{W}^{2})L(m_{j},M_{W})),  \\
\Gamma_{Z}^{(e)}&=& -\frac{3}{2}(1-s_{W}^{2})(\tilde{B}_{0}(0;m_{j},M_{W})
+M_{W}^{2}L(M_{W},m_{j})),  \\
\Gamma_{Z}^{(f+g)}&=& -s_{W}^{2}m_{j}^{2}L(M_{W},m_{j}),  \\
\Gamma_{Z}^{(h)}&=&\frac{1}{8}(-1+2s_{W}^{2})\frac{m_{j}^{2}}{M_{W}^{2}}
(\tilde{B}_{0}(0;M_{W},m_{j})+M_{W}^{2}L(M_{W},m_{j})),  
\end{eqnarray*}

where $\tilde{B}_{0}$ and $\tilde{B}_{1}$ Green functions are defined and 
evaluated in the Appendix ($m_{j}$ denotes the quark mass and
$s_{W}^{2} \equiv \sin^{2}(\Theta_{W})$). 

$D = F_{1}(q^{2}\gamma^{\mu}P_{L}) = F_{1}(-\slashed{q}q^{\mu}P_{L})$
denotes the contribution to the induced quarks-photon vertex to
the second order in the external momentum $q$ from graphs of fig. 3
(a-h) of the ref. \cite{Inami} (Eq. (B.1)). The form factors $F_{1}$
extracted from the two different Lorentz structures, $q^{2}\gamma^{\mu}P_{L}$
and $-\slashed{q}q^{\mu}P_{L}$, must be equal and this is a check
on the validity of our calculation:

\begin{eqnarray}
D=F_{1}(q^{2}\gamma^{\mu}P_{L})=(a+b)+(c)+(d)+(e)+(f+g)+(h) ,
\end{eqnarray}

\begin{eqnarray*}
(a+b)&=&-\frac{1}{3}M_{W}^{2}(1+\frac{1}{2}\frac{m_{\alpha}^{2}}{M_{W}^{2}})
\tilde{B}'_{1}(0;m_{\alpha},M_{W}), \hspace{90 mm} \\
(c)&=&\frac{4}{3}M_{W}^{2}[\tilde{C}_{1}(0;m_{\alpha},M_{W})+
\tilde{C}_{2}(0;m_{\alpha},M_{W})
+2 \tilde{C}'_{3}(0;m_{\alpha},M_{W})
-m_{\alpha}^{2}\tilde{C}'_{0}(0;m_{\alpha},M_{W})], \\
(d)&=&-\frac{2}{3}m_{\alpha}^{2}[m_{\alpha}^{2}\tilde{C}'_{0}(0;m_{\alpha},M_{W})
-2 \tilde{C}'_{3}(0;m_{\alpha},M_{W})-\tilde{C}_{1}(0;m_{\alpha},M_{W})
-\tilde{C}_{2}(0;m_{\alpha},M_{W})], \\
(e)&=&M_{W}^{2}[-4 \tilde{C}'_{3}(0;M_{W},m_{\alpha})
-\frac{1}{2}\tilde{B}'_{0}(0;m_{\alpha},M_{W})
-2 M_{W}^{2} \tilde{C}'_{0}(0;M_{W},m_{\alpha}) \\
&-&\frac{1}{2} \tilde{C}_{0}(0;M_{W},m_{\alpha}) 
 -2 \tilde{C}_{1}(0;M_{W},m_{\alpha})], \\
(f+g)&=&2 m_{\alpha}^{2}M_{W}^{2}\tilde{C}'_{0}(0;M_{W},m_{\alpha}), \\ 
(h)&=&-2 m_{\alpha}^{2} \tilde{C}'_{3}(0;M_{W},m_{\alpha}), 
\end{eqnarray*}

\begin{eqnarray}
D=F_{1}(-\slashed{q}q^{\mu}P_{L})=\{c\}+\{d\}+\{e\}+\{h\}:
\end{eqnarray}

\begin{eqnarray*}
\{c\} &=& \frac{8}{3}M_{W}^{2}[\tilde{C}_{1}(0;m_{\alpha},M_{W})
+\tilde{C}_{2}(0;m_{\alpha},M_{W})],  \hspace{80 mm} \\
\{d\} &=& \frac{4}{3}m_{\alpha}^{2}[\tilde{C}_{1}(0;m_{\alpha},M_{W})
+\tilde{C}_{2}(0;m_{\alpha},M_{W})], \\
\{e\} &=& M_{W}^{2}[\tilde{C}_{0}(0;M_{W},m_{\alpha})
+4 \tilde{C}_{2}(0;M_{W},m_{\alpha})
+4 \tilde{C}_{1}(0;M_{W},m_{\alpha})], \\
\{h\} &=& m_{\alpha}^{2}[2 \tilde{C}_{2}(0;M_{W},m_{\alpha})
+2 \tilde{C}_{1}(0;M_{W},m_{\alpha})
+\frac{1}{2} \tilde{C}_{0}(0;M_{W},m_{\alpha})], \\
\end{eqnarray*}

The definitions and the explicit forms of the Green functions
$\tilde{B}'_{1},\ \tilde{C}_{1},\ \tilde{C}_{2},\ \tilde{C}'_{0},\ 
\tilde{C}'_{3},\ \tilde{B}'_{0},\ \tilde{C}_{0}$ can be found in the 
Appendix.

Only (a-d) graphs of the fig. 3 of ref. \cite{Inami} contribute to
the gauge invariant quarks-gluon vertex in the form factor $F_{g}$:

\begin{eqnarray}
E = F_{g}(q^{2}\gamma^{\mu}P_{L})=F_{g}(-\slashed{q}q^{\mu}P_{L}),
\end{eqnarray}

\begin{eqnarray*}
F_{g}(q^{2}\gamma^{\mu}P_{L})=
2 M_{W}^{2}\{\frac{1}{2}(1+\frac{1}{2}\frac{m_{\alpha}^{2}}{M_{W}^{2}})
\tilde{B}_{1}'(0;m_{\alpha},M_{W})+\tilde{C}_{2}(0;m_{\alpha},M_{W})
+\tilde{C}_{1}(0;m_{\alpha},M_{W}) \\
+2 \tilde{C}'_{3}(0;m_{\alpha},M_{W})
-m_{\alpha}^{2} \tilde{C}'_{0}(0;m_{\alpha},M_{W})
-\frac{1}{2}\frac{m_{\alpha}^{2}}{M_{W}^{2}}
[m_{\alpha}^{2} \tilde{C}'_{0}(0;m_{\alpha},M_{W})
-2 \tilde{C}'_{3}(0;m_{\alpha},M_{W}) \\
-\tilde{C}_{1}(0;m_{\alpha},M_{W})
-\tilde{C}_{2}(0;m_{\alpha},M_{W})]\},
\end{eqnarray*}

\begin{eqnarray*}
F_{g}(-\slashed{q}q^{\mu}P_{L})=
 2 M_{W}^{2}(2+\frac{m_{\alpha}^{2}}{M_{W}^{2}})[\tilde{C}_{1}(0;m_{\alpha},M_{W})
+\tilde{C}_{2}(0;m_{\alpha},M_{W})].
\end{eqnarray*}

Knowing $B_{1},B_{2},C,D$ functions, we construct the gauge 
invariant linear combinations $X,Y,Z$ \cite{Buchalla}:

\begin{eqnarray*}
X = C -4 B_{1},\ Y=C -B_{2},\ Z= C+\frac{1}{4}D,
\end{eqnarray*}

which appear in the amplitudes of the suppressed electroweak
processes \cite{Buras3}. The $B_{1}$ and $B_{2}$ functions are almost equal for
small lepton masses.

Numerical evaluations and conclusions are given in the next
chapter.

\section{Conclusions}

We can accomplish the aim of this paper to verify our
SM results with the standard formulas (see the Appendix) and then to 
compare all the results with the 
functions of the BY theory
containing the Lorentz and gauge invariant cut-off $\Lambda$.

\epsfig{figure=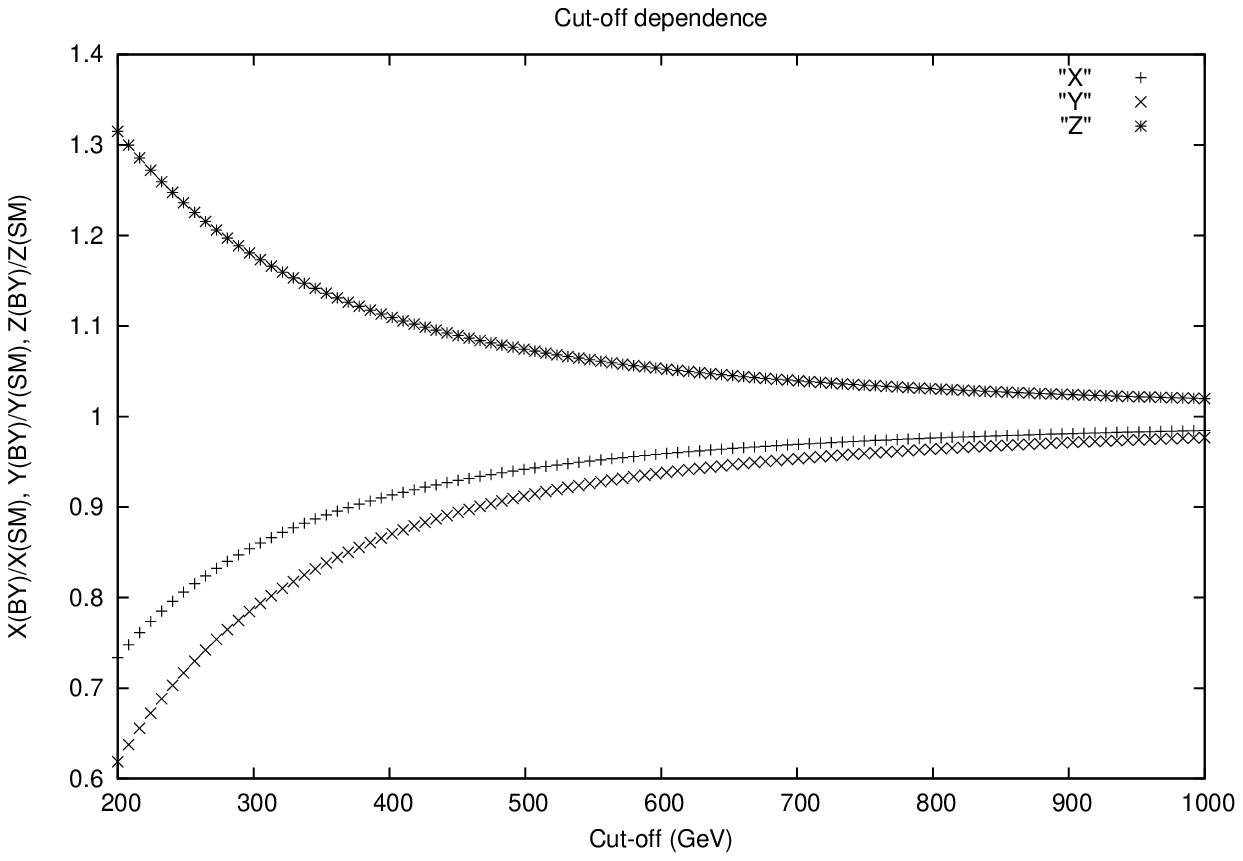, height=90 mm, width=130 mm}

\vspace{25mm}

{\bf Fig. 1: Cut-off ($\Lambda$) dependence of the quotients
$X^{BY}/X^{SM}$, $Y^{BY}/Y^{SM}$ and $Z^{BY}/Z^{SM}$; X,Y,Z
are defined as $X=X(x_{t})-X(x_{u})$
with parameters $m_{u}= 3 MeV$, $m_{t}= 172 GeV$,
$M_{W}= 80.4 GeV$.
}
\newline

We verified that for any function $F=\{B_{1},B_{2},C,D,E\}$ the
following relation is valid $\lim_{\Lambda \rightarrow \infty} F^{BY}(\Lambda)
 = F^{SM}$.

\begin{table}
\caption{Gauge invariant X,Y,Z,E functions evaluated with
$m_{u}=3 MeV$,$m_{c}=1.3 GeV$, $M_{W}=80.4 GeV$, $\Lambda=326 GeV$ and
$\sin^{2}\Theta_{W}=0.23$.}
\begin{tabular}{| c | c | c | c | c | c |} \hline \hline
$\Delta X^{SM}(x_{u},x_{c})$ & $\Delta X^{BY}(x_{u},x_{c})$ & $\Delta X(x_{u},x_{c})$ &
$[\frac{\Delta X(x_{u},x_{c})^{SM}}{\Delta X(x_{u},x_{c})}]^{2}$ &
$[\frac{\Delta X(x_{u},x_{c})^{BY}}{\Delta X(x_{u},x_{c})^{SM}}]^{2}$ &
$[\frac{\Delta X(x_{u},x_{c})^{BY}}{\Delta X(x_{u},x_{c})}]^{2}$
\\  \hline 
$1.5545\times 10^{-3}$ & $1.5487\times 10^{-3}$ & $1.5528\times 10^{-3}$ & 
1.0022 & 0.9925 & 0.9947 \\  \hline \hline
$\Delta Y^{SM}(x_{u},x_{c})$ & $\Delta Y^{BY}(x_{u},x_{c})$ & $\Delta Y(x_{u},x_{c})$ &
$[\frac{\Delta Y(x_{u},x_{c})^{SM}}{\Delta Y(x_{u},x_{c})}]^{2}$ &
$[\frac{\Delta Y(x_{u},x_{c})^{BY}}{\Delta Y(x_{u},x_{c})^{SM}}]^{2}$ &
$[\frac{\Delta Y(x_{u},x_{c})^{BY}}{\Delta Y(x_{u},x_{c})}]^{2}$
\\  \hline 
$1.3244\times 10^{-4}$ & $1.2695\times 10^{-4}$ & $1.3053\times 10^{-4}$ & 
1.0294 & 0.9188 & 0.9458 \\  \hline \hline
$\Delta Z^{SM}(x_{u},x_{c})$ & $\Delta Z^{BY}(x_{u},x_{c})$ & $\Delta Z(x_{u},x_{c})$ &
$[\frac{\Delta Z(x_{u},x_{c})^{SM}}{\Delta Z(x_{u},x_{c})}]^{2}$ &
$[\frac{\Delta Z(x_{u},x_{c})^{BY}}{\Delta Z(x_{u},x_{c})^{SM}}]^{2}$ &
$[\frac{\Delta Z(x_{u},x_{c})^{BY}}{\Delta Z(x_{u},x_{c})}]^{2}$
\\  \hline 
 -1.3492  & -1.3492   & -1.3496 & 
0.9994 & 1.0000 & 0.9995 \\  \hline \hline
$\Delta E^{SM}(x_{u},x_{c})$ & $\Delta E^{BY}(x_{u},x_{c})$ & $\Delta E(x_{u},x_{c})$ &
$[\frac{\Delta E(x_{u},x_{c})^{SM}}{\Delta E(x_{u},x_{c})}]^{2}$ &
$[\frac{\Delta E(x_{u},x_{c})^{BY}}{\Delta E(x_{u},x_{c})^{SM}}]^{2}$ &
$[\frac{\Delta E(x_{u},x_{c})^{BY}}{\Delta E(x_{u},x_{c})}]^{2}$ 
 \\  \hline
 -8.0931  & -8.0931   & -8.0950 & 
0.9995 & 1.0000 & 0.9995 
\\  \hline \hline
\end{tabular}
\end{table}

\begin{table}
\caption{Gauge invariant X,Y,Z,E functions evaluated with
$m_{u}=3 MeV$,$m_{t}=172 GeV$, $M_{W}=80.4 GeV$, $\Lambda=326 GeV$ and
$\sin^{2}\Theta_{W}=0.23$.}
\begin{tabular}{| c | c | c | c | c | c |} \hline \hline
$\Delta X^{SM}(x_{u},x_{t})$ & $\Delta X^{BY}(x_{u},x_{t})$ & $\Delta X(x_{u},x_{t})$ &
$[\frac{\Delta X(x_{u},x_{t})^{SM}}{\Delta X(x_{u},x_{t})}]^{2}$ &
$[\frac{\Delta X(x_{u},x_{t})^{BY}}{\Delta X(x_{u},x_{t})^{SM}}]^{2}$ &
$[\frac{\Delta X(x_{u},x_{t})^{BY}}{\Delta X(x_{u},x_{t})}]^{2}$
\\  \hline 
1.6111 & 1.4098 & 1.5777 & 
1.0428 & 0.7657 & 0.7984 \\  \hline \hline
$\Delta Y^{SM}(x_{u},x_{t})$ & $\Delta Y^{BY}(x_{u},x_{t})$ & $\Delta Y(x_{u},x_{t})$ &
$[\frac{\Delta Y(x_{u},x_{t})^{SM}}{\Delta Y(x_{u},x_{t})}]^{2}$ &
$[\frac{\Delta Y(x_{u},x_{t})^{BY}}{\Delta Y(x_{u},x_{t})^{SM}}]^{2}$ &
$[\frac{\Delta Y(x_{u},x_{t})^{BY}}{\Delta Y(x_{u},x_{t})}]^{2}$
\\  \hline 
1.0595 & 0.8632 & 1.0261 & 
1.0661 & 0.6637 & 0.7076 \\  \hline \hline
$\Delta Z^{SM}(x_{u},x_{t})$ & $\Delta Z^{BY}(x_{u},x_{t})$ & $\Delta Z(x_{u},x_{t})$ &
$[\frac{\Delta Z(x_{u},x_{t})^{SM}}{\Delta Z(x_{u},x_{t})}]^{2}$ &
$[\frac{\Delta Z(x_{u},x_{t})^{BY}}{\Delta Z(x_{u},x_{t})^{SM}}]^{2}$ &
$[\frac{\Delta Z(x_{u},x_{t})^{BY}}{\Delta Z(x_{u},x_{t})}]^{2}$
\\  \hline 
 -1.2658  & -1.4630   & -1.5473 & 
0.6693 & 1.3358 & 0.8940 \\  \hline \hline
$\Delta E^{SM}(x_{u},x_{t})$ & $\Delta E^{BY}(x_{u},x_{t})$ & $\Delta E(x_{u},x_{t})$ &
$[\frac{\Delta E(x_{u},x_{t})^{SM}}{\Delta E(x_{u},x_{t})}]^{2}$ &
$[\frac{\Delta E(x_{u},x_{t})^{BY}}{\Delta E(x_{u},x_{t})^{SM}}]^{2}$ &
$[\frac{\Delta E(x_{u},x_{t})^{BY}}{\Delta E(x_{u},x_{t})}]^{2}$ 
 \\  \hline
 -12.7506  & -12.8194   & -13.3355 & 
0.9142 & 1.0108 & 0.9241 
\\  \hline \hline
\end{tabular}
\end{table}

The following definition is introduced in tables I and II:
$\Delta X(x_{u},x_{t})\equiv X(x_{t})-X(x_{u})$, 
$x_{t}\equiv (\frac{m_{t}}{M_{W}})^{2}$ and
similarly for other functions because these combinations appear
in the amplitudes after applying the unitarity of the quark mixing
matrix \cite{Inami}.
The functions without superscripts
are the standard ones \cite{Buchalla}, and those with $SM$ superscripts are
the SM functions evaluated in this paper, whereas those with $BY$ superscripts
are the functions attributed to the $BY$ theory with $\Lambda$.
One can notice that the functions for the c-quark
agree for our SM outputs and the standard functions \cite{Buchalla,Inami}
(table I),
but there are disagreements for the t-quark for $D$ and $E$ functions (table II).
We check our SM formulas for $D$ and $E$ through current conservation,
hence these standard functions \cite{Inami} are not accurate for the
heaviest t-quark. The vertex contributions for $\Gamma^{Z}$ agree perfectly
graph by graph with the Inami-Lim \cite{Inami} (A.1)-(A.3) results,
but their sum is an approximation
\cite{Palle6}.

As expected, a comparison between the SM ($\Lambda = \infty$) and
the BY theory  ($\Lambda = 326 GeV$) functions reveals a difference of
roughly 10 - 30 \%, thus opening the possibility to discover the deviations
from the SM for the suppressed electroweak processes and meson mixings,
provided that
the hadron matrix elements can be calculated with a sufficient
precision \cite{Carrasco} and the measurements have small systematic 
and statistical errors \cite{LHCb}.
The most recent result of the LHCb \cite{LHCb2}, measuring the branching
fraction of $B_{s}^{0} \rightarrow \phi \mu^{+}\mu^{-}$, appears
to be lower than the SM prediction, thus favouring our theory of noncontractible
space \cite{Palle1}.
\newline
\newline

{\bf Appendix}
\newline
\newline

For the purpose of comparison with our results,
we give the standard functions widely used in the literature \cite{Buchalla} 

\begin{eqnarray*}
B(x)&=&\frac{1}{4}[\frac{x}{1-x}+\frac{x \ln x}{(x-1)^{2}}], \\
C(x)&=&\frac{x}{8}[\frac{x-6}{x-1}+\frac{3x+2}{(x-1)^2}\ln x], \\
D(x)&=&-\frac{4}{9}\ln x+\frac{-19x^{3}+25x^{2}}{36(x-1)^{3}}
+\frac{x^{2}(5x^{2}-2x-6)}{18(x-1)^4}\ln x, \\
E(x)&=&-\frac{2}{3}\ln x+\frac{x^{2}(15-16x+4x^{2})}{6(1-x)^{4}}
\ln x+\frac{x(18-11x-x^{2})}{12(1-x)^{3}}.
\end{eqnarray*}

Integrals appearing in $B_{1},B_{2},C$ look like:

\begin{eqnarray*}
g_{1}^{SM;BY}(m_{1},m_{2},m_{3}) &\equiv & \int^{\infty;\Lambda}_{0}
dx x [x+m_{1}^{2}]^{-1}[x+m_{2}^{2}]^{-1}[x+m_{3}^{2}]^{-2},
\\
f_{1}^{SM;BY}(m_{1},m_{2},m_{3}) &\equiv &\int^{\infty;\Lambda}_{0}
dx x^{2} [x+m_{1}^{2}]^{-1}[x+m_{2}^{2}]^{-1}[x+m_{3}^{2}]^{-2}, \\
L^{SM;BY}(m_{1},m_{2}) &\equiv &-2 \int^{\infty;\Lambda}_{0}
dq q^{3} [q^{2}+m_{1}^{2}]^{-2}[q^{2}+m_{2}^{2}]^{-1},
\end{eqnarray*}

and the explicit forms for the SM are:

\begin{eqnarray*}
g_{1}^{SM}(m_{1},m_{2},m_{3})=(-(m_{1}^{2}(m_{2}^{2}-m_{3}^{2})^{2}
\ln m_{1}^{2})+m_{2}^{2}(m_{1}^{2}-m_{3}^{2})^{2} \ln m_{2}^{2}-
(m_{1}^{2}-m_{2}^{2}) \\
\times ((m_{1}^{2}-m_{3}^{2})(m_{2}^{2}-m_{3}^{2})
+(m_{1}^{2}m_{2}^{2}-m_{3}^{4})\ln m_{3}^{2}))/
((m_{1}^{2}-m_{2}^{2})(m_{1}^{2}-m_{3}^{2})^{2}(m_{2}^{2}-m_{3}^{2})^{2}), \\
f_{1}^{SM}(m_{1},m_{2},m_{3})=(m_{1}^{4}(m_{2}^{2}-m_{3}^{2})^{2}
\ln m_{1}^{2}-m_{2}^{4}(m_{1}^{2}-m_{3}^{2})^{2}\ln m_{2}^{2}
+(m_{2}^{2}-m_{1}^{2})m_{3}^{2} \\
\times ((m_{1}^{2}-m_{3}^{2})(m_{3}^{2}-m_{2}^{2})+
(-2m_{1}^{2}m_{2}^{2}+(m_{1}^{2}+m_{2}^{2})m_{3}^{2})\ln m_{3}^{2}))/  \\
((m_{1}^{2}-m_{2}^{2})(m_{1}^{2}-m_{3}^{2})^{2}(m_{2}^{2}-m_{3}^{2})^{2}), 
\end{eqnarray*}
\begin{eqnarray*}
L^{SM}(m_{1},m_{2})&=&\frac{1}{m_{2}^{2}}(\frac{m_{1}^{2}}{m_{2}^{2}}
-1)^{-2}(1-\frac{m_{1}^{2}}{m_{2}^{2}}+\ln \frac{m_{1}^{2}}{m_{2}^{2}}), \\
L^{BY}(m_{1},m_{2})&=&-(\Lambda^{2}(m_{1}^{2}-m_{2}^{2})+
m_{2}^{2}(\Lambda^{2}+m_{1}^{2})(\ln\frac{m_{2}^{2}}{m_{1}^{2}}
+\ln \frac{\Lambda^{2}+m_{1}^{2}}{\Lambda^{2}+m_{2}^{2}}))/
((\Lambda^{2}+m_{1}^{2})(m_{1}^{2}-m_{2}^{2})^{2}).
\end{eqnarray*}

We do not show 
the analogous lengthy analytic expressions for other integrals with the cut-off.
The analytic forms are checked with the numerically evaluated integrals
to arbitrary precision. 

The following definitions of the real parts of Green functions have been employed
\newline
($\frac{\partial \tilde{B}(k^{2})}{\partial k^{2}}\equiv \tilde{B}'(k^{2})$):

\begin{eqnarray*}
\frac{\imath}{16\pi^{2}}\tilde{B}_{0}(k^{2};m_{1},m_{2})&\equiv &
(2\pi)^{-4}\int d^{4}q(q^{2}-m_{1}^{2})^{-1}((q+k)^{2}-m_{2}^{2})^{-1},
  \hspace{30 mm}\\
\frac{\imath}{16\pi^{2}}k_{\mu}\tilde{B}_{1}(k^{2};m_{1},m_{2})&\equiv &
(2\pi)^{-4}\int d^{4}q q_{\mu}(q^{2}-m_{1}^{2})^{-1}
((q+k)^{2}-m_{2}^{2})^{-1},  \\
\tilde{B}_{1}(0;m_{1},m_{2})&=&\frac{1}{2}(-\tilde{B}_{0}(0,m_{1},m_{2})
+(m_{2}^{2}-m_{1}^{2}) \tilde{B}_{0}'
(0,m_{1},m_{2})), \\
\tilde{B}_{1}'(0;m_{1},m_{2})&=& \frac{1}{2}(-\tilde{B}_{0}'(0;m_{1},m_{2})
+\frac{1}{2}(m_{2}^{2}-m_{1}^{2})\tilde{B}_{0}''(0;m_{1},m_{2})).
\end{eqnarray*}

The real parts of these Green functions, denoted by the superscripts
SM and BY for the evaluation without and with the UV cut-off $\Lambda$
respectively, turn out to be:

\begin{eqnarray*}
\tilde{B}_{0}^{SM}(0;m_{1},m_{2})&=&\Delta_{UV}-x\ln x/(x-1),\ 
x=\frac{m_{1}^{2}}{ m_{2}^{2}},\ 
\Delta_{UV}\equiv UV\ infinity,  \\
 \tilde{B}_{0}^{'SM}(0;m_{1},m_{2}) &=&
\frac{1}{2}\frac{m_{1}^{2}+m_{2}^{2}}{(m_{1}^{2}-m_{2}^{2})^{2}}
-\frac{m_{1}^{2}m_{2}^{2}}{(m_{2}^{2}-m_{1}^{2})^{3}}
\ln \frac{m_{2}^{2}}{m_{1}^{2}}, \\
\tilde{B}_{0}^{''SM}(0;m_{1},m_{2})&=&[(m_{1}^{2}-m_{2}^{2})
(m_{1}^{4}+10 m_{1}^{2}m_{2}^{2}+m_{2}^{4})
+6 m_{1}^{2}m_{2}^{2}(m_{1}^{2}+m_{2}^{2})\ln \frac{m_{2}^{2}}
{m_{1}^{2}}]/[3(m_{1}^{2}-m_{2}^{2})^{5}],
\end{eqnarray*}

\begin{eqnarray*}
\tilde{B}_{0}^{BY}(0;m_{1},m_{2})&=&\int^{\Lambda^{2}}_{0}
dy \frac{y}{(y+m_{1}^{2})(y+m_{2}^{2})} \\
&=& (m_{1}^{2}\ln\frac{\Lambda^{2}
+m_{1}^{2}}{m_{1}^{2}}-m_{2}^{2}\ln\frac{\Lambda^{2}+m_{2}^{2}}
{m_{2}^{2}})/(m_{1}^{2}-m_{2}^{2}), 
\end{eqnarray*}

\begin{eqnarray*}
\frac{\partial \tilde{B}_{0}^{BY}}{\partial k^{2}}(0;m_{1},m_{2})&=&
\frac{1}{2}(\frac{\partial \hat{B}_{0}^{BY}}
{\partial k^{2}}(0;m_{1},m_{2})+
\frac{\partial \hat{B}_{0}^{BY}}
{\partial k^{2}}(0;m_{2},m_{1})),\hspace{90 mm} 
\end{eqnarray*}

\begin{eqnarray*}
\frac{\partial \hat{B}_{0}^{BY}}
{\partial k^{2}}(0;m_{1},m_{2})&=&
m_{2}^{2}\int^{\Lambda^{2}}_{0}dy \frac{y}{(y+m_{1}^{2})
(y+m_{2}^{2})^{3}} \hspace{100 mm}  \hspace{50 mm} \\
&=& [\Lambda^{2}(-2 m_{1}^{2}m_{2}^{4}+2 m_{2}^{6}+
\Lambda^{2}(m_{2}^{4}-m_{1}^{4}))+
2 m_{1}^{2}m_{2}^{2}(\Lambda^{2}+m_{2}^{2})^{2} \\
&\times & (\ln \frac{m_{1}^{2}}{m_{2}^{2}}-\ln \frac{\Lambda^{2}+
m_{1}^{2}}{\Lambda^{2}+m_{2}^{2}})]
/[2 (\Lambda^{2}+m_{2}^{2})^{2}(m_{2}^{2}-m_{1}^{2})^{3}], 
\end{eqnarray*}

\begin{eqnarray*}
\frac{\partial^{2} \tilde{B}_{0}^{BY}}{\partial (k^{2})^{2}}(0;m_{1},m_{2})&=&
\frac{1}{2}(\frac{\partial^{2} \hat{B}_{0}^{BY}}{\partial (k^{2})^{2}}(0;m_{1},m_{2})
+\frac{\partial^{2} \hat{B}_{0}^{BY}}{\partial (k^{2})^{2}}(0;m_{2},m_{1})), \\
\frac{\partial^{2} \hat{B}_{0}^{BY}}{\partial (k^{2})^{2}}(0;m_{1},m_{2})
&=&2 m_{2}^{2}\int^{\Lambda^{2}}_{0}d y \frac{y (m_{2}^{2}-y)}
{(y+m_{1}^{2})(y+m_{2}^{2})^{5}}.
\end{eqnarray*}

Let us finally define and evaluate the necessary vertex Green
functions:

\begin{eqnarray*}
\frac{\imath}{16\pi^{2}}\tilde{C}_{0}(k^{2};m_{1},m_{2})\equiv 
(2\pi)^{-4}\int d^{4}q[(q+k)^{2}-m_{1}^{2}]^{-1}[q^{2}-m_{1}^{2}]^{-1}
 [(q+k/2)^{2}-m_{2}^{2}]^{-1}, \\
\frac{\imath}{16\pi^{2}}k_{\mu}\tilde{C}_{1}(k^{2};m_{1},m_{2})\equiv 
(2\pi)^{-4}\int d^{4}q q_{\mu}[(q+k)^{2}-m_{1}^{2}]^{-1}[q^{2}-m_{1}^{2}]^{-1}
 [(q+k/2)^{2}-m_{2}^{2}]^{-1}, \\
\frac{\imath}{16\pi^{2}}[k_{\mu}k_{\nu}\tilde{C}_{2}(k^{2};m_{1},m_{2})
+g_{\mu\nu}\tilde{C}_{3}(k^{2};m_{1},m_{2})]  \hspace{60 mm} \\
=(2\pi)^{-4}\int d^{4}q q_{\mu}q_{\nu}
[(q+k)^{2}-m_{1}^{2}]^{-1}[q^{2}-m_{1}^{2}]^{-1}
 [(q+k/2)^{2}-m_{2}^{2}]^{-1}.
\end{eqnarray*}

It is easy to verify that $\tilde{C}_{1}(k^{2};m_{1},m_{2})=
-\frac{1}{2} \tilde{C}_{0}(k^{2};m_{1},m_{2})$ and:

\begin{eqnarray*}
\tilde{C}_{2}(k^{2};m_{1},m_{2})=\frac{1}{3 k^{2}}[
\tilde{B}_{0}(k^{2}/4;m_{1},m_{2})
+2 \tilde{B}_{1}(k^{2}/4;m_{1},m_{2})
-(m_{1}^{2}-k^{2}) \tilde{C}_{0}(k^{2};m_{1},m_{2})], \\
\tilde{C}_{3}(k^{2};m_{1},m_{2})=\frac{1}{3}[
\frac{1}{2}\tilde{B}_{0}(k^{2}/4;m_{1},m_{2})
-\frac{1}{2}\tilde{B}_{1}(k^{2}/4;m_{1},m_{2})
+(m_{1}^{2}-k^{2}/4) \tilde{C}_{0}(k^{2};m_{1},m_{2})].
\end{eqnarray*}

We need Green functions evaluated at $k^{2}=0$, thus the following
values of $\tilde{C}_{0}$ and $\tilde{C}'_{0}$ should be calculated:

\begin{eqnarray*}
\tilde{C}_{0}^{SM;BY}(0;m_{1},m_{2}) &=& -2 \int_{0}^{\infty;\Lambda}
d q q^{3}(q^{2}+m_{1}^{2})^{-2}(q^{2}+m_{2}^{2})^{-1}, \\
\tilde{C}_{0}^{'SM;BY}(0;m_{1},m_{2}) &=& -\frac{1}{2} \int_{0}^{\infty;\Lambda}
d q q^{3}(2 m_{1}^{2}+q^{2})(q^{2}+m_{1}^{2})^{-4}(q^{2}+m_{2}^{2})^{-1}.
\end{eqnarray*}

Straightforward integration yields:

\begin{eqnarray*}
\tilde{C}_{0}^{SM}(0;m_{1},m_{2}) &=& (m_{2}^{2}-m_{1}^{2}+m_{2}^{2}
\ln \frac{m_{1}^{2}}{m_{2}^{2}})/(m_{1}^{2}-m_{2}^{2})^{2}, \\
\tilde{C}_{0}^{'SM}(0;m_{1},m_{2}) &=&-[5 m_{1}^{6}
-9 m_{1}^{2}m_{2}^{4}+ 4 m_{2}^{6} - 6 m_{1}^{2}m_{2}^{2}(2m_{1}^{2}-
m_{2}^{2})\ln \frac{m_{1}^{2}}{m_{2}^{2}}]/[24 m_{1}^{2}(m_{1}^{2}
-m_{2}^{2})^{4}], \\
\tilde{C}_{0}^{BY}(0;m_{1},m_{2}) &=& - [(m_{1}^{2}-m_{2}^{2}+
m_{2}^{2} \ln \frac{m_{2}^{2}}{m_{1}^{2}})/(m_{1}^{2}-m_{2}^{2})^{2} 
+(-m_{1}^{4}+m_{1}^{2}m_{2}^{2}+(\Lambda^{2}+m_{1}^{2})m_{2}^{2} \\
&\times&\ln (\Lambda^{2}+m_{1}^{2})-(\Lambda^{2}+m_{1}^{2})m_{2}^{2}
\ln (\Lambda^{2}+m_{2}^{2}))/((\Lambda^{2}+m_{1}^{2})
(m_{1}^{2}-m_{2}^{2})^{2})].
\end{eqnarray*}

Lengthy analytical expressions for integrals in
$\frac{\partial^{2} \hat{B}_{0}^{BY}}{\partial (k^{2})^{2}}(0;m_{1},m_{2})$ 
and $\tilde{C}_{0}^{'BY}(0;m_{1},m_{2})$ are not given.
Displayed functions suffice to fix $\tilde{C}_{2}$ appearing
in the amplitudes:

\begin{eqnarray*}
\tilde{C}_{2}(k^{2};m_{1},m_{2})&=&\frac{a_{-1}}{k^{2}}+a_{0}+{\cal O}(k^{2}), \\
\tilde{C}_{2}(0;m_{1},m_{2}) \equiv a_{0} &=& \frac{1}{3}
[\frac{1}{4}\tilde{B}_{0}'(0;m_{1},m_{2})
+\frac{1}{2} \tilde{B}_{1}'(0;m_{1},m_{2})
+\tilde{C}_{0}(0;m_{1},m_{2}) \\
&-&m_{1}^{2}\tilde{C}_{0}'(0;m_{1},m_{2})].
\end{eqnarray*}

\end{document}